\documentclass[11pt]{article}
\usepackage[a4paper,margin=2.6cm]{geometry}
\usepackage{amsmath,amssymb,amsthm}
\usepackage{float}
\usepackage{longtable,booktabs,array}
\usepackage{url}
\usepackage{hyperref}

\theoremstyle{definition}

\theoremstyle{remark}

\newcommand{\Q}{\mathbb Q}

\newcommand{\cA}{\mathcal A}
\newcommand{\Fbar}{\overline{F}}

\newcommand{\fd}{\mathfrak d}

\newcommand{\arcsec}{\operatorname{arcsec}}

\title{Parallel Integration over Simple Radical Extensions\\ in Mixed Towers: Charlwood's Integrals}
\author{Sam Blake}
\date{\today}

\input{charlwood_numbers}

\begin{document}
\maketitle

\begin{abstract}
Charlwood's 2008 paper \emph{Integration on computer algebra systems} lists fifty indefinite integrals of real elementary functions --- ten worked in detail and forty in an appendix --- chosen to defeat Maple, Mathematica and the TI-89 of the day: products of inverse functions, arcsines and logarithms of algebraic functions, radicals over trigonometric functions. We run all fifty through the SymPy implementation of the parallel (Risch--Norman) method of Part II of this series, in the form in which they are written, with the tower of the integrand built automatically, and verify every answer by differentiation. The method returns \npm\ verified integrals of the $50$ and no wrong answer; the one it does not return, $\int\arcsin(x\sqrt{1-x^2})\,dx$, is not elementary --- the holomorphic-remainder certificate of Part II shows this after one integration by parts --- and in its own tower, a curve of genus three, the implementation withholds the certificate because the one hypothesis it cannot verify there is the completeness of the $S'$-units over a special prime of the curve variable. The degree bounds of Part II at every place --- the exponents of the special primes and the degrees in the variables below the top --- are proved on every rung for thirty-eight of the fifty integrals and cut the ans\"atze to a third of the size the classical guess gives them, without changing a result or the running time. The residues at a prime of degree eight that one of the integrals needs are computed in the residue field, by the residue polynomial of Rothstein and Trager inside Algorithm~2 of Part II, and each residue class is realised from its sheet polynomial by Hensel lifting and linear algebra. We give the solutions in the coordinates of the towers, where the mechanisms of Part II --- $S'$-units over the special primes, Pell units of conics and of a genus-two curve, torsion on elliptic curves, a Risch differential equation over a curve --- are visible, and we compare, on the same machine with the same time limit and the same verification, with the Risch--Trager--Bronstein integrator of FriCAS (\nfr\ of $50$, typically in a tenth of a second) and of AXIOM (\nax\ of $50$, with three wrong answers). The comparison is an assessment of a prototype against a mature implementation: the parallel method is slower by a factor of six on the median integral and by more than an order of magnitude on the worst cases, and the profile shows exactly where --- in the nonlinear norm search for the $S'$-units, in the primitive elements of nested number fields and in the retry ladder of guessed inputs, everything else having been moved into one number field per tower --- and what to change. Nothing in the integrator is tuned to the suite: every step it takes is a step of Part II, and every failure is reported as such.
\end{abstract}

\section{Introduction}\label{sec:intro}

In 2008 Charlwood published a short paper \cite{Charlwood08} whose subject is the gap between what a student can integrate by hand and what a computer algebra system integrates unaided. Ten indefinite integrals of real elementary functions are worked in detail --- a product of two inverse functions, an arcsine of a difference of square roots, a logarithm of an algebraic function, trigonometric integrands whose rationalisation is far from obvious --- and forty more are listed in an appendix. Maple 11, Mathematica 6.0.2 and the TI-89 were tested on the ten; the score-card in the paper is four, seven and four successes respectively, several of the ``successes'' being answers in terms of elliptic integrals or complex-valued expressions that a calculus student could not use. Charlwood's point was pedagogical: when a system fails, a substitution or an integration by parts found by hand produces a new integral the system can handle. The paper has since become a standard small benchmark for integration software.

Part II of this series \cite{PartII} developed the parallel (Risch--Norman) method for towers containing one simple radical anywhere in the tower, and Charlwood's integrals are exactly the kind of input that theory was written for: the derivative of $\arcsin(x)$ introduces the curve $y^2=1-x^2$ below the primitive $\arcsin(x)$; $\log(1+x\sqrt{1+x^2})$ is a logarithm of an element of the function field of $y^2=x^2+1$; $\tan(x)\,\sqrt{1+\tan^4(x)}$ is a radical over the hypertangent $t=\tan(x)$. The present paper is a companion to Part II, with four aims.
\begin{enumerate}
\item To run all fifty of Charlwood's integrals through the SymPy implementation of Part II, in the form in which a user would type them: the tower is built automatically from the surface expression by the rules of Section~\ref{sec:tower}, the integral is computed by the pipeline of Part II, and every returned integral is verified by differentiation.
\item To give the solutions the method finds, in the coordinates of the tower where that is more illuminating than the surface form, and for every failure to say exactly which step of the pipeline is responsible and why.
\item To compare, on the same machine and with the same time limit, with the Risch--Trager--Bronstein integrator of FriCAS \cite{FriCAS} and of AXIOM \cite{AXIOM} --- the reference implementations of the recursive algorithm for mixed towers \cite{Bronstein90} --- and to verify their answers by the same protocol.
\item To assess the state of the implementation: which steps of the published algorithm it covers, where its limits are, where its time goes, and what the next pieces of work are. Every failure is reported with the step of the pipeline responsible, every choice of policy in the retry ladder is stated as such, and nothing in the integrator is tuned to the suite.
\end{enumerate}

The outcome is summarised in Table~\ref{tab:summary}. The parallel method returns \npm\ verified integrals of the fifty, in \totpm\ seconds in all; FriCAS returns \nfr\ (and leaves the one non-elementary integral unevaluated), AXIOM \nax. The one integral the parallel method does not return, A39, is not elementary, and its certificate is the one the implementation cannot yet produce inside the tower.

\emph{Related work.} The benchmark itself is \cite{Charlwood08}; parts of it reappear in the test suites of Rubi \cite{Rubi} and in Parisse's comparison of Giac with other systems. The theory is Parts I and II of this series \cite{PartI,PartII}, whose notation we use without further comment: $L=K_0(y)(t_{j+1},\dots,t_n)$ with $y^2=q$, the working ring $\cA$, the denominator divisor $\fd_D$, the valuation shift $\delta_P=1+v_P(\fd_D)$ at normal primes, special primes, the residues $\tau_P(f)$ and canonical residues $\hat\tau_P(f)$, $S$-units and the realisation of residue divisors, and the three exits of the algorithm --- an integral, a certificate of non-elementarity, or ``failed''.

\section{The implementation under test}\label{sec:tower}

\subsection{From a surface expression to a tower}

Part II works with an integrand given as a coordinate pair $(f_0,f_1)\leftrightarrow f_0+f_1y$ over an explicitly described tower. A benchmark of fifty integrals written the way Charlwood wrote them needs the tower to be built automatically; the module \texttt{build\_tower.py} does this by a small set of rules applied innermost first, until the integrand is a rational function of the generators and $y$. The rules are the obvious ones, but the order and the normalisations matter --- a sign decided at the wrong place, or a root flattened over a coefficient that is not constant, changes the tower and with it the integral --- so we state them.

\begin{itemize}
\item[(R1)] \emph{Trigonometric functions of one argument $a$} are handled as a family through a tangent, never through an exponential. If the integrand is odd in $\sin(a)$ (resp.\ $\cos(a)$) and $a$ is the integration variable, the variable is changed to $u=\cos(a)$ (resp.\ $\sin(a)$); if it is even in $(\sin(a),\cos(a))$, the generator is $t=\tan(a)$ with $Dt=(1+t^2)Da$; otherwise $t=\tan(a/2)$. Radicands and the arguments of opaque functions such as $\log(\sin(x))$ are examined by the same parity rules through placeholders.
\item[(R2)] $\log(a)$, $\arctan(a)$, $\operatorname{artanh}(a)$ become primitives with rational derivatives; $\exp(a)$, $\tan(a)$, $\tanh(a)$ become hyper\-exponential and hyper\-tangent monomials. $\arcsin(a)$, $\arccos(a)$, $\operatorname{arsinh}(a)$, $\operatorname{arcosh}(a)$, $\arcsec(a)$, \dots\ become primitives whose derivative goes through the radical $\sqrt{1\mp a^2}$ (or $\sqrt{a^2-1}$), which becomes the radical of the tower if none exists.
\item[(R3)] A root $b^{k/m}$ whose base is $cg+r$, with $g$ a generator, $c$ a constant and $r$ free of $g$ and of $y$, is \emph{flattened} (Lemma~3.2 of Part II): $g\mapsto(u^m-r)/c$. Every other root with $m=2$ is the radical $y^2=b$, after $b$ has been normalised to a squarefree polynomial by $\sqrt{N/D}=\sqrt{ND}/|D|$ and $\sqrt{s^2b'}=|s|\sqrt{b'}$, the signs $|D|$, $|s|$ being decided at a rational sample point of the real domain of the integrand. A factor of the squarefree radicand that is itself a generator $g$, positive at the sample point, is split off as the pure root $\sqrt g$, which then flattens: $\sqrt{2t(1+t^2)}=\sqrt2\,\sqrt t\,\sqrt{1+t^2}$ in A27, $\sqrt{u(1-u)}=\sqrt u\,\sqrt{1-u}$ in Problem~9.
\item[(R4)] If a second, unflattenable radical is demanded and the existing one is a conic $y^2=ag^2+bg+c$, the conic is parametrised: by Euler's substitution $w=y+\sqrt a\,g$ when $a$ is a square, else through a rational point $(g_0,y_0)$ found by a small search, $w=(y-y_0)/(g-g_0)$; the parameter is oriented so that $w>0$ at the sample point. The tower is then transcendental over $\Q(w)$ and the new radical takes its place.
\end{itemize}

\subsection{The integrator}\label{sec:integrator}

The integrator (\texttt{parallel\_mixed.py}) runs Algorithm~4 of Part II with its retry ladder: the base run with the tower specials over $\Q$, the special exponents and degree bounds proved by Algorithm~6 of Part II where its criteria apply and, where they do not, guessed and raised in turn, as (I5) below describes; for a conic radicand, the parametrisation (R4) before any work over $\Fbar$; the specials split over $\Fbar$; and for a quartic radicand with a square leading coefficient, the change to the cubic model. Before the ladder, a tower that has become a single generator $w$ over the curve with $Dw=r(w)\neq1$ --- a flattened root or the parameter of (R4) --- is rescaled to $d/dw$ (Lemma~3.4 of Part II), so that it is the setting of Part I, exact bounds and certificate included; in the suite this concerns Problem~7 alone. Each rung is a guessed input or a change of model in the sense of Remark~9.1 of Part II. Five points of the implementation bear directly on the results below; the third and the fifth are Algorithms~5 and~6 of Part II, the others are not spelled out there.

\begin{itemize}
\item[(I1)] \emph{Canonical residues in mixed towers.} At a normal prime of Hermite order the residue is the canonical residue of Proposition~7.6 of Part II, computed by the local reduction of Algorithm~2 --- $\mu\gets\overline{\pi^{k+\delta}f}$, $c\gets$ a lift of $-\mu/(k\lambda_P)$, $f\gets f-D(c\pi^{-k})$ --- in which $c$ is a function of the upper generators and $D$ is the derivation of the whole tower: when a primitive sits above the curve, $D$ does not act trivially on the coefficients of the completion, and the reduction of the one-generator case, a Laurent expansion with constant coefficients, does not apply. The loop is implemented (\texttt{\_canonical\_residues}) at unramified normal primes with $\delta=1$ and constant coordinates, the leading class $\mu$ being read from the local expansion of $y$ about the place; a transcendental tower ($q$ absent) realises a residue divisor over such a prime by the principal primes $g-\rho$.
\item[(I2)] \emph{Realisation over several primes.} A residue divisor may be supported over several irreducible $p_i\in R_0$ and be realised by a function whose norm $c\prod_ip_i^{k_i}$ is not a power of any one of them. Algorithm~3 is therefore run over the product of the primes carrying the divisor (\texttt{\_realise\_joint}) before the ``needs torsion'' exit; the coefficients are checked to be constant on the hits and to cover every place with a nonzero residue. On a cubic model the torsion step (Algorithm~3(d)) likewise acts on the whole leftover divisor --- its places summed by the group law, the order of the sum being the order of the class --- rather than on the classes $[P-\infty]$ of the single places, which need not be torsion when the divisor is.
\item[(I3)] \emph{Assembly and solution of the linear system over one number field.} The system is linear in the unknown coefficients, and every constant that enters it --- the coefficients of the $Dt_i$, of $q$ and of the residual, the norms of the $S'$-units, the roots of a special split over $\Fbar$ --- is carried as an element of one number field $K=\Q(\theta)$ whose primitive element is computed once per system (Algorithm~5 of Part II). With $\Delta$ the least common multiple of the denominators of the $Dt_i$, every column $D(m/D_v)$ and $D(my/D_v)$, $m$ a monomial, is a pair of polynomials of $K[t]$ over the one denominator $D_v^2\Delta\cdot2q$, computed by the product rule in the sparse polynomial ring with no cancellation; the few remaining columns (units, $S'$-units, specials) are cancelled once; each coordinate is brought over the least common multiple of its denominators, which is exact in $K[t]$ whatever the constants; the coefficient of each monomial is a row of a sparse matrix over $K$, the matrix is reduced to row echelon form, the solution with the free unknowns set to zero is checked on the matrix, and the rational part of the answer is cancelled in $K[t]$ and written back in the radicals of $\Q$. No cancellation or simplification touches an expression that contains an unknown, and the field is never re-derived. The equations are those of Part II up to a polynomial factor per coordinate, and the assembly and the solve are together a small fraction of the running time (Section~\ref{sec:assessment}); in the first version of the implementation, which formed the columns as expressions and checked the solution by substitution and simplification, they were three quarters of it. The same field is the field of the tower (Remark~9.6 of Part II): the coordinate arithmetic of the pairs, the realisation of residue classes (I4) and the torsion arithmetic of Algorithm~3(d) run in it, with exact zero tests; and the once-per-integrand part of Algorithm~4 --- classification, residues and their realisation, tower specials, units and $S'$-units, the residual --- is computed once and reused by every rung of the retry ladder, which re-enters at the ansatz.
\item[(I4)] \emph{Residues at primes of any degree.} At a prime $p$ of degree more than four in the curve variable $g$ --- where explicit roots would carry nested radicals of high degree --- the residues are computed in the residue field $K[g]/(p)$ without roots. With $\tau=e_P\,(fp/Dp)$ written as a pair $\tau_0+\tau_1y$ and reduced modulo $p$, the residues on the two sheets over a root $\alpha$ are $\tau_0(\alpha)\pm\tau_1(\alpha)y_0$, and their minimal polynomial over $K$,
\[
R(z)=\operatorname{res}_g\bigl(p,\ (z-\tau_0)^2-\tau_1^2q\bigr),
\]
the residue polynomial of Rothstein and Trager, gives the residue values $c$ and the fields $K(c)$ through its irreducible factors; constancy is the statement that $\tau_0,\tau_1 \bmod p$ are free of the other generators. The places with residue $c$ lie over $p_c=\gcd\bigl(p,\ \tau_1^2q-(c-\tau_0)^2\bigr)$ on the sheet $y=Y_0(g):=(c-\tau_0)/\tau_1 \bmod p_c$, a polynomial with $Y_0^2\equiv q\pmod{p_c}$, so a residue class is a pair (prime, sheet polynomial) rather than a list of points; where $\tau_1$ vanishes both sheets carry $c$ and the divisor over that factor is principal. The class is realised by linear algebra over $K(c)$: $Y_0$ is lifted by Hensel's iteration to $Y$ with $Y^2\equiv q\pmod{p_c^m}$, the condition that $u=a+by$ vanish to order $m$ along the class is the linear condition $a+bY\equiv0\pmod{p_c^m}$, and the pole orders $(n_+,n_-)$ at the two places at infinity are fixed in turn, so that $N(u)$ has degree exactly $m\deg p_c$ and is a constant times $p_c^m$; the least $m\le12$ with a solution gives the logand, with coefficient $c/m$ (\texttt{\_residue\_classes}, \texttt{\_realise\_class}). Classes of equal value over several primes are realised jointly, as in (I2). The bound on $m$ is a guessed input in the sense of Remark~9.1; a class of larger order is a torsion question for the division polynomials of Parts I and III.
\item[(I5)] \emph{Degree bounds at every place, and the certificate.} The exponent of every special prime and the degree bound in every generator are computed by Algorithm~6 of Part II from the shift of the derivation at the places over the special and over $t_i=\infty$ (Theorem~8.7 and Corollary~8.9 of Part II), and each is recorded as \emph{proved} when one of the criteria (K1)--(K4) of Proposition~8.11 applies at every place over it, or as a \emph{guess} --- the classical bound of \cite[\S10.3]{Bronstein05}, raised by the retry count --- when none does (\texttt{\_place\_bounds}; the verbose trace prints the criterion at every place). The retry over the special exponents runs only while a guess is in play: a rung on which every bound is proved is tried once. One kind of place is left to the guess by the implementation although the theory decides it: a special prime over the curve variable that is a branch prime, or at which the two coordinates of a derivative have equal valuation, so that the Gauss valuation on the pair is only a lower bound and the expansions of Part I, \S6.1 would be needed. The criterion (K1) is applied in its extended form, with attaining hyperexponential generators: the kernel degrees vanish unless $\rho_v$ is a $\Q$-linear combination of their constants $\mu_k$, which the implementation decides, when $\rho_v$ and the $\mu_k$ are numbers, by linear algebra over $\Q$ in a common number field (\texttt{\_in\_qspan}); Problem~10 is the one integral of the suite where this arises. The certificate of Proposition~9.2(b) is issued exactly when every bound in force is proved, the unit group is known, the specials are irreducible over $\Fbar$ (split, or nothing to split), no special lies over the curve variable --- there the $S'$-units come from the bounded norm search of (L2) and are not known to be complete --- and the residual is \emph{verified} residue-free by re-running the residue analysis on it; in every other case the exit is ``failed''. On the suite, thirty-eight integrals are computed with every bound proved on every rung and twelve with a guess in play; against the classical guess with the same ladder, the ans\"atze summed over the rungs shrink from $7053$ unknowns to $1966$ and the rungs from $103$ to $74$, with no change in any result and none in the running time, the linear algebra being about a second of the \totpm\,s (Section~\ref{sec:assessment}).
\end{itemize}

\subsection{Protocol}\label{sec:protocol}

All fifty integrals were run with the script \texttt{charlwood.py} accompanying this paper. Each integral is given at most $300$ seconds of wall-clock time; timings are for a single core of a 16-core Apple workstation, SymPy 1.14 under Python 3.11. A returned integral $I$ is accepted only if $I'-f$ evaluates to less than $10^{-18}$ at three rational points of the real domain of $f$ at $30$ digits; a status tuple is recorded verbatim. The same integrals were given to FriCAS 1.3.6 and AXIOM (August 2014) in one fresh process each, with the same limit (\texttt{cas\_compare.py}); the answer was captured in input form and re-verified by the same test in SymPy, so that all three systems are judged by one criterion, independent of their own numerical evaluators (which matters: FriCAS's own \texttt{complexNumeric} evaluates its answer to Problem~8 incorrectly, although the answer is right). The times reported for FriCAS and AXIOM are those printed by \texttt{)set messages time on} for the \texttt{integrate} call alone; the times for the parallel method include the construction of the tower.

% generated by charlwood_report.py
\begingroup\small\setlength{\tabcolsep}{4pt}\setlength{\LTcapwidth}{\textwidth}
\begin{longtable}{|l|>{$}l<{$}|c@{\ }r|c@{\ }r|c@{\ }r|}
\caption{The fifty integrals of \cite{Charlwood08}: outcome and time in seconds for the parallel method (PM), FriCAS 1.3.6 and AXIOM (2014); Charlwood's $\theta$ is written $x$. $\checkmark$ an integral verified by differentiation, \textsf{NE} a certificate of non-elementarity, \textsf{F} ``failed'', \textsf{T} the 300-second limit, \textsf{TE} the tower could not be built, \textsf{W} a returned answer that fails the differentiation test, \textsf{U} an unevaluated integral in the answer, \textsf{E} an error. The last rows count the outcomes and give the total, the mean and the median of the times shown.}\label{tab:summary}\\
\hline
\multicolumn{1}{|c|}{} & \multicolumn{1}{c|}{integrand} & \multicolumn{2}{c|}{PM} & \multicolumn{2}{c|}{FriCAS} & \multicolumn{2}{c|}{AXIOM}\\ \hline
\endfirsthead
\multicolumn{8}{l}{Table~\ref{tab:summary} (continued)}\\[2pt]
\hline
\multicolumn{1}{|c|}{} & \multicolumn{1}{c|}{integrand} & \multicolumn{2}{c|}{PM} & \multicolumn{2}{c|}{FriCAS} & \multicolumn{2}{c|}{AXIOM}\\ \hline
\endhead
\hline\endfoot
\endlastfoot
P1 & \arcsin(x)\ln(x) & $\checkmark$ & 0.5 & $\checkmark$ & 0.1 & $\checkmark$ & 0.2\\
P2 & x\arcsin(x)/\sqrt{1-x^2} & $\checkmark$ & 0.2 & $\checkmark$ & 0.1 & $\checkmark$ & 0.1\\
P3 & \arcsin(\sqrt{x+1}-\sqrt x) & $\checkmark$ & 2.0 & $\checkmark$ & 52.8 & $\textsf{E}$ & --\\
P4 & \ln(1+x\sqrt{1+x^2}) & $\checkmark$ & 6.2 & $\checkmark$ & 0.1 & $\checkmark$ & 0.1\\
P5 & \cos^2(x)/\sqrt{\cos^4(x)+\cos^2(x)+1} & $\checkmark$ & 2.6 & $\checkmark$ & 0.2 & $\checkmark$ & 0.5\\
P6 & \tan(x)\sqrt{1+\tan^4(x)} & $\checkmark$ & 0.6 & $\checkmark$ & 0.2 & $\checkmark$ & 0.5\\
P7 & \tan(x)/\sqrt{\sec^3(x)+1} & $\checkmark$ & 0.2 & $\checkmark$ & 0.2 & $\checkmark$ & 0.4\\
P8 & \sqrt{\tan^2(x)+2\tan(x)+2} & $\checkmark$ & 7.0 & $\checkmark$ & 0.7 & $\checkmark$ & 12.2\\
P9 & \sin(x)\arctan(\sqrt{\sec(x)-1}) & $\checkmark$ & 0.5 & $\checkmark$ & 0.1 & $\textsf{E}$ & --\\
P10 & x^3e^{\arcsin(x)}/\sqrt{1-x^2} & $\checkmark$ & 0.5 & $\checkmark$ & 0.1 & $\checkmark$ & 0.1\\
\hline
A1 & x\ln(1+x^2)\ln(x+\sqrt{1+x^2})/\sqrt{1+x^2} & $\checkmark$ & 1.3 & $\checkmark$ & 0.1 & $\checkmark$ & 0.1\\
A2 & \arctan(x+\sqrt{1-x^2}) & $\checkmark$ & 3.6 & $\checkmark$ & 0.1 & $\textsf{W}$ & 0.2\\
A3 & x\arctan(x+\sqrt{1-x^2})/\sqrt{1-x^2} & $\checkmark$ & 3.1 & $\checkmark$ & 0.1 & $\textsf{W}$ & 0.2\\
A4 & \arcsin(x)/(1+\sqrt{1-x^2}) & $\checkmark$ & 0.7 & $\checkmark$ & 0.1 & $\checkmark$ & 0.1\\
A5 & \ln(x+\sqrt{1+x^2})/(1-x^2)^{3/2} & $\checkmark$ & 1.1 & $\checkmark$ & 0.1 & $\textsf{E}$ & --\\
A6 & \arcsin(x)/(1+x^2)^{3/2} & $\checkmark$ & 1.7 & $\checkmark$ & 0.1 & $\textsf{E}$ & --\\
A7 & \ln(x+\sqrt{x^2-1})/(1+x^2)^{3/2} & $\checkmark$ & 1.1 & $\checkmark$ & 0.1 & $\textsf{E}$ & --\\
A8 & \ln(x)/(x^2\sqrt{x^2-1}) & $\checkmark$ & 0.3 & $\checkmark$ & 0.1 & $\checkmark$ & 0.1\\
A9 & \sqrt{1+x^3}/x & $\checkmark$ & 0.2 & $\checkmark$ & 0.1 & $\checkmark$ & 0.1\\
A10 & x\ln(x+\sqrt{x^2-1})/\sqrt{x^2-1} & $\checkmark$ & 0.1 & $\checkmark$ & 0.1 & $\checkmark$ & 0.1\\
A11 & x^3\arcsin(x)/\sqrt{1-x^4} & $\checkmark$ & 1.3 & $\checkmark$ & 0.1 & $\textsf{E}$ & --\\
A12 & x^3\operatorname{arcsec}(x)/\sqrt{x^4-1} & $\checkmark$ & 1.6 & $\checkmark$ & 0.1 & $\textsf{E}$ & --\\
A13 & x\arctan(x)\ln(x+\sqrt{1+x^2})/\sqrt{1+x^2} & $\checkmark$ & 0.4 & $\checkmark$ & 0.1 & $\checkmark$ & 0.1\\
A14 & x\ln(1+\sqrt{1-x^2})/\sqrt{1-x^2} & $\checkmark$ & 0.3 & $\checkmark$ & 0.1 & $\checkmark$ & 0.1\\
A15 & x\ln(x+\sqrt{1+x^2})/\sqrt{1+x^2} & $\checkmark$ & 0.1 & $\checkmark$ & 0.1 & $\checkmark$ & 0.1\\
A16 & x\ln(x+\sqrt{1-x^2})/\sqrt{1-x^2} & $\checkmark$ & 1.4 & $\checkmark$ & 0.1 & $\checkmark$ & 0.1\\
A17 & \ln(x)/(x^2\sqrt{1-x^2}) & $\checkmark$ & 0.3 & $\checkmark$ & 0.1 & $\checkmark$ & 0.1\\
A18 & x\arctan(x)/\sqrt{1+x^2} & $\checkmark$ & 0.2 & $\checkmark$ & 0.1 & $\checkmark$ & 0.1\\
A19 & \arctan(x)/(x^2\sqrt{1-x^2}) & $\checkmark$ & 1.1 & $\checkmark$ & 0.1 & $\checkmark$ & 0.2\\
A20 & x\arctan(x)/\sqrt{1-x^2} & $\checkmark$ & 0.6 & $\checkmark$ & 0.1 & $\checkmark$ & 0.1\\
A21 & \arctan(x)/(x^2\sqrt{1+x^2}) & $\checkmark$ & 0.7 & $\checkmark$ & 0.1 & $\checkmark$ & 0.2\\
A22 & \arcsin(x)/(x^2\sqrt{1-x^2}) & $\checkmark$ & 0.4 & $\checkmark$ & 0.1 & $\checkmark$ & 0.1\\
A23 & x\ln(x)/\sqrt{x^2-1} & $\checkmark$ & 0.3 & $\checkmark$ & 0.1 & $\checkmark$ & 0.1\\
A24 & \ln(x)/(x^2\sqrt{1+x^2}) & $\checkmark$ & 0.2 & $\checkmark$ & 0.1 & $\checkmark$ & 0.1\\
A25 & x\operatorname{arcsec}(x)/\sqrt{x^2-1} & $\checkmark$ & 0.3 & $\checkmark$ & 0.1 & $\textsf{E}$ & --\\
A26 & x\ln(x)/\sqrt{1+x^2} & $\checkmark$ & 0.2 & $\checkmark$ & 0.1 & $\checkmark$ & 0.1\\
A27 & \sqrt{\sin(x)}/(1+\sin^2(x)) & $\checkmark$ & 4.2 & $\checkmark$ & 0.1 & $\textsf{E}$ & --\\
A28 & (1+x^2)/((1-x^2)\sqrt{1+x^4}) & $\checkmark$ & 3.7 & $\checkmark$ & 0.1 & $\checkmark$ & 0.1\\
A29 & (1-x^2)/((1+x^2)\sqrt{1+x^4}) & $\checkmark$ & 2.9 & $\checkmark$ & 0.1 & $\checkmark$ & 0.1\\
A30 & \ln(\sin(x))/(1+\sin(x)) & $\checkmark$ & 0.5 & $\checkmark$ & 0.1 & $\checkmark$ & 0.2\\
A31 & \ln(\sin(x))\sqrt{1+\sin(x)} & $\checkmark$ & 0.9 & $\checkmark$ & 0.1 & $\textsf{E}$ & --\\
A32 & \sec(x)/\sqrt{\sec^4(x)-1} & $\checkmark$ & 0.6 & $\checkmark$ & 0.2 & $\textsf{W}$ & 0.2\\
A33 & \tan(x)/\sqrt{1+\tan^4(x)} & $\checkmark$ & 0.5 & $\checkmark$ & 0.1 & $\checkmark$ & 0.4\\
A34 & \sin(x)/\sqrt{1-\sin^6(x)} & $\checkmark$ & 0.7 & $\checkmark$ & 0.2 & $\checkmark$ & 0.3\\
A35 & \sqrt{\sqrt{\sec(x)+1}-\sqrt{\sec(x)-1}} & $\checkmark$ & 4.1 & $\checkmark$ & 7.7 & $\textsf{E}$ & --\\
A36 & x\ln(x^2+1)\arctan^2(x) & $\checkmark$ & 0.1 & $\checkmark$ & 0.1 & $\checkmark$ & 0.1\\
A37 & \arctan(x\sqrt{1+x^2}) & $\checkmark$ & 2.1 & $\checkmark$ & 0.1 & $\checkmark$ & 0.1\\
A38 & \arctan(\sqrt{x+1}-\sqrt x) & $\checkmark$ & 0.4 & $\checkmark$ & 0.1 & $\checkmark$ & 0.1\\
A39 & \arcsin(x\sqrt{1-x^2}) & $\textsf{F}$ & 1.6 & $\textsf{U}$ & 0.1 & $\textsf{E}$ & --\\
A40 & \arctan(x\sqrt{1-x^2}) & $\checkmark$ & 2.8 & $\checkmark$ & 0.2 & $\checkmark$ & 0.2\\
\hline
\multicolumn{2}{|l|}{$\checkmark$ verified integrals} & \multicolumn{2}{c|}{49} & \multicolumn{2}{c|}{49} & \multicolumn{2}{c|}{35}\\
\multicolumn{2}{|l|}{$\textsf{F}$ ``failed''} & \multicolumn{2}{c|}{1} & \multicolumn{2}{c|}{0} & \multicolumn{2}{c|}{0}\\
\multicolumn{2}{|l|}{$\textsf{T}$ time limit reached} & \multicolumn{2}{c|}{0} & \multicolumn{2}{c|}{0} & \multicolumn{2}{c|}{0}\\
\multicolumn{2}{|l|}{$\textsf{W}$ wrong answers} & \multicolumn{2}{c|}{0} & \multicolumn{2}{c|}{0} & \multicolumn{2}{c|}{3}\\
\multicolumn{2}{|l|}{$\textsf{U}$ unevaluated integrals} & \multicolumn{2}{c|}{0} & \multicolumn{2}{c|}{1} & \multicolumn{2}{c|}{0}\\
\multicolumn{2}{|l|}{$\textsf{E}$ errors} & \multicolumn{2}{c|}{0} & \multicolumn{2}{c|}{0} & \multicolumn{2}{c|}{12}\\
\multicolumn{2}{|l|}{total time (s)} & \multicolumn{2}{c|}{68.2} & \multicolumn{2}{c|}{66.6} & \multicolumn{2}{c|}{18.3}\\
\multicolumn{2}{|l|}{mean time (s)} & \multicolumn{2}{c|}{1.4} & \multicolumn{2}{c|}{1.3} & \multicolumn{2}{c|}{0.5}\\
\multicolumn{2}{|l|}{median time (s)} & \multicolumn{2}{c|}{0.6} & \multicolumn{2}{c|}{0.1} & \multicolumn{2}{c|}{0.1}\\
\hline
\end{longtable}
\endgroup

\section{The ten problems}\label{sec:problems}

For each problem we give the tower built by (R1)--(R4), the mechanism of Part II that carries the integral, the solution in the coordinates of the tower (with the substitutions that return it to $x$), the time, and Charlwood's own answer where it differs in form. Every displayed formula is checked by differentiation by the script \texttt{charlwood\_paper\_check.py}.

\subsection*{Problem 1: $\int\arcsin(x)\,\ln(x)\,dx$}
Tower $x$, $t_1=\log(x)$, $t_2=\arcsin(x)$ over $y^2=1-x^2$, with $Dt_1=1/x$ and $Dt_2=y/(1-x^2)$; integrand $t_1t_2$. The prime $(x)$ is special (the argument of a logarithm), and the $S'$-units over it found by the norm search are $1\pm y$ (norm $x^2$) and $-1-ix\pm y$ (norm $2ix$). Every bound is proved --- the exponent $0$ of the special and the degree $1$ in $x$ by (K1), the degrees $2$ in $t_1$ and $t_2$ by (K2) --- and the ansatz, of $42$ unknowns, is solved at the first attempt, in \tpm{P1}\,s:
\[
\int\arcsin(x)\ln(x)\,dx=x\,t_1t_2-x\,t_2+(t_1-2)\,y+t_1-\log(y-1),
\]
i.e.\ Charlwood's $-\ln(\sqrt{1-x^2}-1)+\ln(x)+(x\arcsin(x)+\sqrt{1-x^2})\ln(x)-x\arcsin(x)-2\sqrt{1-x^2}$. The logarithm $\log(1-\sqrt{1-x^2})$ that every system has to find is here an $S'$-unit of the curve over the special prime $x$: its divisor is supported at the two places over $x=0$ and at infinity, and it is invisible to every residue --- exactly the situation of Remark~7.8 of Part II. Maple~11 could not do the original integral in 2008 (Maple~8 could not even do the second integral after the parts step); FriCAS returns the same answer with $\tfrac12\log(y+1)-\tfrac12\log(y-1)$ in place of our $-\log(y-1)+\log(x)$ (the two differ by $\tfrac12\log(y^2-1)=\log(x)+$const).

\subsection*{Problem 2: $\int x\arcsin(x)/\sqrt{1-x^2}\,dx$}
Tower $x$, $t=\arcsin(x)$ over $y^2=1-x^2$; the integrand is $-tx/(x^2-1)\cdot y=tx/y$. Both branch primes carry a pole of order $1$, which is sub-critical ($\delta=2$), so there are no residues and no Hermite part, and the ansatz ($13$ unknowns) returns $x-y\,t$ in \tpm{P2}\,s. This is the archetype Charlwood describes --- ``an inverse trigonometric function that we would differentiate in the parts process'' --- and in the parallel method there is no parts step: the $t$-linear block of the system is the integration by parts.

\subsection*{Problem 3: $\int\arcsin(\sqrt{x+1}-\sqrt x)\,dx$}
This is the nested-radical example, and the tower construction is the whole story. $\sqrt x$ is a pure root and flattens to $u=\sqrt x$ (R3); $\sqrt{x+1}=\sqrt{u^2+1}$ is a conic with square leading coefficient, and when the arcsine demands a further radical the conic is parametrised by Euler's substitution $w=u+\sqrt{u^2+1}=\sqrt x+\sqrt{x+1}$ (R4). In the parameter, $\sqrt{x+1}-\sqrt x=1/w$, so $t=\arcsin(1/w)$ is a primitive whose derivative needs $\sqrt{1-1/w^2}=\sqrt{w^2-1}/w$: the final tower is $w$, $t$ over the conic $y^2=w^2-1$, with $Dw=2w^3/(w^4-1)$ and $Dt=-2w^2y/((w^2-1)(w^4-1))$, and the integrand is $-t$. The prime $(w)$ is special (it divides the moving denominator), with $S'$-units $\mp w-i\pm y$ (norm $\pm2iw$); its exponent $2$ and the degree $4$ in $w$ are proved by (K1) and the degree $2$ in $t$ by (K2), so the system ($36$ unknowns) is solved at the first attempt, in \tpm{P3}\,s in all, two thirds of which is the verification of the nested-radical answer by differentiation:
\[
\int\arcsin(\sqrt{x+1}-\sqrt x)\,dx=\frac{(2w^2+1)\sqrt{w^2-1}}{8w^2}+\frac{2w^4-w^2+2}{8w^2}\,\arcsin\Bigl(\frac1w\Bigr) .
\]
Since $w^2+w^{-2}=(\sqrt x+\sqrt{x+1})^2+(\sqrt{x+1}-\sqrt x)^2=4x+2$, the coefficient of the arcsine is $x+\tfrac38$, which is Charlwood's; his algebraic part $\tfrac{\sqrt2}4\sqrt{\sqrt{x^2+x}+x}+\tfrac{\sqrt2}8\sqrt{\sqrt x(\sqrt{x+1}-\sqrt x)^3}$ is the first term above (both are $\sqrt{2\sqrt x\,w}\,(2w^2+1)/(8w^2)$). Charlwood's solution needed two rationalising substitutions and an integration by parts found by a student; Mathematica~6 gave a complex-domain answer and the TI-89 nothing. FriCAS finds the same answer, but in \tfr{P3}\,s: the two square roots are handled there as an algebraic extension of degree four, whereas flattening and the Euler parametrisation reduce the problem to genus zero before the integrator starts.

\subsection*{Problem 4: $\int\ln(1+x\sqrt{1+x^2})\,dx$}
Tower $x$, $t=\log(1+xy)$ over $y^2=x^2+1$, integrand $t$. The norm of the logand is $N(1+xy)=1-x^2(x^2+1)=-(x^4+x^2-1)$, so the moving denominator of $Dt$ is $x^4+x^2-1=(x^2+\phi)(x^2-1/\phi)$, $\phi=(1+\sqrt5)/2$, and this quartic is the special prime. The $S'$-units over it have coefficients in $\Q(\sqrt{(1-2i)/5})$. The exponent of the quartic is a guess in the $x$-tower --- the two coordinates of $Dt$ have equal valuation there, the case (I5) leaves to the guess --- and with the exponents $0,1,2$ the base run fails three times ($18$, $42$, $66$ unknowns); the conic is then parametrised by $w=(y-1)/x$, the specials in the $w$-tower are $w\pm1$ and the quartic $w^4+2w^3-2w^2+2w+1$, every bound there is proved by (K1) and (K2) --- exponent $1$ at $w\pm1$, $0$ at the quartic, degree $2$ in $w$ and in $t$ --- so the tower is tried once ($12$ unknowns), and only after the quartic is split over $\Fbar$ into its four linear factors is the system solved ($15$ unknowns; \tpm{P4}\,s in all). The answer, in Charlwood's own notation --- his $t$ is our Euler parameter $w$, his $a=\phi$ ---
\begin{align*}
\int\ln(1+x\sqrt{1+x^2})\,dx&=x\ln(1+x\sqrt{1+x^2})-2x+2\sqrt a\arctan\Bigl(\sqrt a\,w-\frac1{\sqrt a}\Bigr)\\
&\qquad+\frac1{\sqrt a}\ln\Bigl(\frac{w+a-\sqrt a}{w+a+\sqrt a}\Bigr),\qquad
w=\frac{\sqrt{x^2+1}-1}x,\quad a=\frac{1+\sqrt5}2 ,
\end{align*}
is what the pipeline returns once its conjugate complex logarithms are paired. Two remarks. The needed logands are the linear factors $w-\rho_i$ of the split quartic, which the theory offers only after the specials are split over $\Fbar$ (Theorem~6.1 of Part II); in the original $x$-tower the same logands would be $S'$-units over the four factors $x\pm\sqrt{\pm\phi^{\mp1}}$ of the quartic, which the search bounded to norms $c\,p^k$ with $\deg b\le1$ does not produce. And the order of the ladder --- parametrise before splitting --- is a policy, discussed in Section~\ref{sec:assessment}. FriCAS (\tfr{P4}\,s) returns an answer of the same shape with the same constants $\sqrt{\sqrt5-1}$; Charlwood reports that Mathematica~6 needed $\tanh^{-1}$ and Maple~11 gave a more complicated result.

\subsection*{Problem 5: $\int\cos^2(x)/\sqrt{\cos^4(x)+\cos^2(x)+1}\,dx$}
Even in $(\sin,\cos)$, so $t=\tan(x)$ (R1), and the radicand becomes $\cos^4(x)\,(1+\sec^2(x)+\sec^4(x))=(t^4+3t^2+3)/(1+t^2)^2$: the tower is $x$, $t$ over the quartic $y^2=t^4+3t^2+3$, of genus one, and the integrand is $y/(t^4+3t^2+3)=1/y$. Charlwood remarks that ``it might appear at first that this is an elliptic integral''; it is an integral \emph{on} an elliptic curve, and the mechanism that makes it elementary is torsion. The branch primes are sub-critical, there are no residues, and the only special is the hypertangent's $t^2+1$; the $S'$-units over it (norms $t^2+1$ and $-3(t^2+1)$) do not suffice, nor do those over $t\pm i$; every bound is proved on both rungs --- the exponent $0$ of the special and the degree $0$ in $t$ by (K1), the degree $1$ in $x$ by (K2) --- so each is tried once ($10$ and $11$ unknowns), and after the two failures the ladder changes the model: with $s=y-t^2-\tfrac32$ the curve becomes $y^2=-8s^3-24s^2+6s$, the specials are $2s-1$ and $2s+3$, and the $S'$-units over them are the Miller functions of the torsion classes $[P-\infty]$ of order $3$ at $(\tfrac12,\pm2i)$ and of order $6$ at $(-\tfrac32,\pm6i)$. The system ($10$ unknowns) is solved, \tpm{P5}\,s in all, and the answer in the original variables is
\[
\int\frac{\cos^2(x)\,dx}{\sqrt{\cos^4(x)+\cos^2(x)+1}}=-x-\frac23\arctan\Bigl(\frac{3y-3t^2-5}{2ty-2t^3-3t}\Bigr),\qquad
\begin{aligned}t&=\tan(x),\\ y&=\sqrt{t^4+3t^2+3},\end{aligned}
\]
which differs from Charlwood's $-\tfrac13\arcsin(\cos^3(x))$ by a constant on each interval. Mathematica~6 and Maple~11 answered in elliptic $F$ and $\Pi$; FriCAS returns, in \tfr{P5}\,s,
\[
\frac16\arctan\Bigl(\frac{2\cos^3(x)\sin(x)\sqrt{\cos^4(x)+\cos^2(x)+1}}{2\cos^6(x)-1}\Bigr).
\]

\subsection*{Problem 6: $\int\tan(x)\sqrt{1+\tan^4(x)}\,dx$}
Tower $x$, $t=\tan(x)$ over $y^2=t^4+1$ (genus one), integrand $ty$. The special $t^2+1$ has the $S'$-units $-t^2-1+\sqrt2\pm y$ and $t^2+1+\sqrt2\pm y$, of norms $(2\mp2\sqrt2)(t^2+1)$; with them and the unit $t^2+y$ (norm $-1$) the first attempt succeeds ($18$ unknowns, \tpm{P6}\,s):
\begin{align*}
\int\tan(x)\sqrt{1+\tan^4(x)}\,dx&=\frac y2-\frac{\sqrt2+1}2\log(y+t^2)+\frac{\sqrt2}2\log\Bigl(\frac{\sqrt2-1-t^2-y}{\sqrt2-1-t^2+y}\Bigr),\\
t&=\tan(x),\qquad y=\sqrt{t^4+1},
\end{align*}
Charlwood's answer with $\sqrt2\ln|\sec(x)|$ and $\ln|\sqrt{2+2\tan^2(x)}-\tan^2(x)+1|$ regrouped. Mathematica~6 answered in elliptic functions.

\subsection*{Problem 7: $\int\tan(x)/\sqrt{\sec^3(x)+1}\,dx$}
The integrand is odd in $\sin(x)$, so the variable is changed to $u=\cos(x)$; then $\sqrt{\sec^3(x)+1}=\sqrt{(1+u^3)/u^3}$, whose normalisation demands $\sqrt u$ --- a pure root, flattened to $v=\sqrt u=\sqrt{\cos(x)}$ with $Dv=1/(2v)$ --- and the remaining radical is $y^2=v^6+1$: a curve of genus \emph{two}. The tower is the single generator $v$ over the curve, and the integrator rescales it to $d/dv$ (Lemma~3.4 of Part II): the integrand $-v/y$ becomes $-2v^2/y$, both branch primes are sub-critical, and the whole integral is carried by the fundamental unit $v^3+y$ ($N=-1$) found by the continued fraction of $y$:
\[
\int\frac{\tan(x)\,dx}{\sqrt{\sec^3(x)+1}}=-\frac23\log\bigl(\cos^{3/2}(x)+\sqrt{\cos^3(x)+1}\bigr)=\frac13\ln\Bigl(\frac{\sqrt{\sec^3(x)+1}-1}{\sqrt{\sec^3(x)+1}+1}\Bigr)+C ,
\]
in \tpm{P7}\,s. It is the same mechanism as Cohen's and Schultz's integrals in Part II, on a hyperelliptic curve, with the trigonometric substitution done by the parity rule rather than by Charlwood's ``multiply and divide by $\sec(x)$''.

\subsection*{Problem 8: $\int\sqrt{\tan^2(x)+2\tan(x)+2}\,dx$}
Tower $x$, $t=\tan(x)$ over the conic $y^2=t^2+2t+2$, integrand $y$. The special $t^2+1$ has $S'$-units with coefficients in $\Q(\sqrt5,\sqrt{\sqrt5-1})$ --- norms $\tfrac{\sqrt5-3}2(t^2+1)$ and $-\tfrac{\sqrt5+3}2(t^2+1)$ --- and the first attempt ($10$ unknowns) succeeds; the \tpm{P8}\,s are, in three nearly equal parts, the norm search --- a nonlinear solve over $\Q$ ---, the construction of three fields of degree sixteen --- primitive elements over nested radicals, one for each set of constants that the $S'$-units and their logarithms bring in --- and the columns and the system over them. The answer is $\log(t+1+y)$ plus four logarithms with complex coefficients that pair into Charlwood's two arctangents with $\sqrt{(\sqrt5\pm1)/2}$ and his logarithm with coefficient $\sqrt{(\sqrt5-1)/8}$. Charlwood solved it by hand through the substitution $z=\tan(\theta/2)$ with $\tan(\theta)=\tan(x)+1$ and a partial-fraction decomposition of a quartic; FriCAS's answer (\tfr{P8}\,s) is three thousand characters long and contains $\sqrt[4]5$ and $\sqrt{(\sqrt5-5)/(\sqrt5-3)}$, and FriCAS's own numerical evaluation of it is wrong although the expression is a correct antiderivative (Section~\ref{sec:protocol}); AXIOM returns a list of two candidate answers, of which the second is correct.

\subsection*{Problem 9: $\int\sin(x)\arctan\bigl(\sqrt{\sec(x)-1}\bigr)\,dx$}
Odd in $\sin(x)$: $u=\cos(x)$, and $\sqrt{\sec(x)-1}=\sqrt{(1-u)/u}$ demands $\sqrt u$, flattened to $v=\sqrt{\cos(x)}$, leaving the conic $y^2=1-v^2$ and the primitive $t=\arctan(y/v)$ with $Dt=y/(2v^3-2v)$. The integrand is $-t$; there are no poles at all, and the ansatz ($19$ unknowns) gives, in \tpm{P9}\,s,
\begin{align*}
\int\sin(x)\arctan\bigl(\sqrt{\sec(x)-1}\bigr)\,dx&=\bigl(\tfrac12-v^2\bigr)t+\frac{v\,y}2\\
&=\bigl(\tfrac12-\cos(x)\bigr)\arctan\bigl(\sqrt{\sec(x)-1}\bigr)+\frac{\sqrt{\cos(x)(1-\cos(x))}}2 ,
\end{align*}
Charlwood's answer with $\operatorname{arcsec}\bigl(\sqrt{\sec(x)}\bigr)=\arctan\bigl(\sqrt{\sec(x)-1}\bigr)$. Maple~11 gave nothing here.

\subsection*{Problem 10: $\int x^3e^{\arcsin(x)}/\sqrt{1-x^2}\,dx$}
Tower $x$, $t_1=\arcsin(x)$, $t_2=e^{t_1}$ over $y^2=1-x^2$: a hyperexponential whose derivative $Dt_2=t_2/y$ goes through the curve --- the setting of Bronstein's \emph{Risch differential equation on an algebraic curve} and of \S10.14 of Part II. The integrand is $x^3t_2/y$; the branch poles are sub-critical, the only special is $(t_2)$, and the $t_2$-linear block of the system ($34$ unknowns) is the Risch equation, solved in \tpm{P10}\,s. Every bound is proved, one of them by the extended form of (K1): at $x=\infty$ the arcsine and the exponential above it both attain the shift of the derivation, with $\lambda_1=\mu=\mp i$ at the two places and $\rho_v=-1$, and since $\rho_v$ is not a rational multiple of $\mu$ the kernel degrees vanish and the degree in $x$ is $3$ (the classical guess is $5$); the exponent of $(t_2)$ and the degrees in $t_1$, $t_2$ are proved by (K2). The answer
\[
\int\frac{x^3e^{\arcsin(x)}}{\sqrt{1-x^2}}\,dx=\frac{e^{\arcsin(x)}}{10}\bigl(x^3+3x-3(x^2+1)\sqrt{1-x^2}\bigr)
\]
is exactly Charlwood's form. Neither Maple~11 nor the TI-89 could do this integral, nor the simpler $\int e^{\arcsin(x)}\,dx$.

\section{The forty integrals of the appendix}\label{sec:appendix}

We group the appendix integrals by the mechanism they exercise. The status and time of each is in Table~\ref{tab:summary}; the solutions are in the file \texttt{charlwood\_results.json} produced by the run (surface form, tower form, and the full trace of the pipeline).

\subsection*{Inverse functions times algebraic functions: A8, A10, A14--A18, A20, A23--A26}
These are the textbook integration-by-parts integrals,
\[
\int\frac{x\ln(x+\sqrt{x^2\pm1})}{\sqrt{x^2\pm1}},\quad \int\frac{x\arctan(x)}{\sqrt{1\pm x^2}},\quad \int\frac{\ln(x)}{x^2\sqrt{1\pm x^2}},\quad \int\frac{x\operatorname{arcsec}(x)}{\sqrt{x^2-1}},\ \dots ,
\]
and all but A16 are solved at the first attempt in under a second, with every bound proved, by the same pattern as Problems~1 and 2: sub-critical branch poles, one special (the argument of the logarithm or arctangent, or the hypertangent's $1+x^2$), and the $S'$-units over it. Two are worth a comment. In A16, $\int x\ln(x+\sqrt{1-x^2})/\sqrt{1-x^2}$, the special is $2x^2-1=N(x+y)$ and the answer needs the logands $\pm\sqrt2/2\pm y$ and $w\pm(1\pm\sqrt2)$ in the Euler parameter, which the ladder reaches only after the conic parametrisation and the $\Fbar$-split (\tpm{A16}\,s). In A18, $\int x\arctan(x)/\sqrt{1+x^2}$, the logand $\log(x+\sqrt{1+x^2})$ that the answer requires is invisible to every residue (it is the Pell unit of the conic, of norm $-1$) and enters as the unit candidate of Remark~7.8 of Part II.

\subsection*{Hermite-order poles under a primitive: A4, A19, A21, A22}
$\int\arcsin(x)/(1+\sqrt{1-x^2})$, $\int\arctan(x)/(x^2\sqrt{1\mp x^2})$, $\int\arcsin(x)/(x^2\sqrt{1-x^2})$ have a pole of order $2$ at the normal prime $(x)$ under a primitive $t$. The Hermite part is $D_v=x$; the residues at the two places over $x=0$ are canonical residues at a pole of Hermite order (Proposition~7.6 of Part II), and must be computed by the local reduction $f\mapsto f-D(c\pi^{-k})$ with $c$ a \emph{function of $t$}, since $D$ does not act trivially on the coefficients of the completion (the remark following that proposition; this is the loop (I1) of Section~\ref{sec:integrator}). The residues at $(0,\pm1)$ are $\pm1$, the logands $1\pm y$ (A21: $-x-1\pm y$ of norm $2x$) are realised by the norm search, and the three integrals are solved in about a second each; for instance
\begin{align*}
\int\frac{\arcsin(x)\,dx}{x^2\sqrt{1-x^2}}&=\log(x)-\frac{\sqrt{1-x^2}\arcsin(x)}x ,\\
\int\frac{\arctan(x)\,dx}{x^2\sqrt{1+x^2}}&=\log\Bigl(\frac{\sqrt{x^2+1}-x-1}{\sqrt{x^2+1}+x+1}\Bigr)+\log(x+\sqrt{x^2+1})-\frac{\sqrt{1+x^2}\arctan(x)}x .
\end{align*}
A4 goes through the conic parametrisation instead, where the pole is at the transcendental prime $(w)$ and the logand $1+w^2$ is a special.

\subsection*{Two conics, and elliptic quartics from the Euler parameter: A5, A6, A7, A11, A12}
$\int\ln(x+\sqrt{1+x^2})/(1-x^2)^{3/2}$ and its relatives contain two radicals. The first --- from the logarithm or the arcsine --- is a conic, parametrised by (R4); the second, $(1\mp x^2)^{3/2}$ or $\sqrt{1-x^4}$, becomes a quartic in the parameter, $w^4+6w^2+1$ or $4w^4\pm24w^2+4$, of genus one, with a pole of order $3$ at its branch primes (a Hermite part at a branch prime, $\delta=2$). The specials are $w$ or $w^2+1$ from the moving denominator, with $S'$-units such as $-2w\pm y$ (norm $-(w^2+1)^2$) and $-2iw^2-2i\pm y$; every bound is proved --- the exponent of the special ($0$, or $2$ at $w$ in A12 and at $w^2+1$ in A11) and the degree $4$ in $w$ by (K1), the degree $2$ in $t$ by (K2) --- and the systems, of $36$--$41$ unknowns, are solved at the first attempt, in \tpm{A5}, \tpm{A6}, \tpm{A7}, \tpm{A11} and \tpm{A12}\,s. The answers are naturally expressed in $w$; A11 for instance is
\begin{align*}
\int\frac{x^3\arcsin(x)}{\sqrt{1-x^4}}dx&=\frac{(w^2-1)\arcsin(x)-w}{2(w^2+1)^2}\,\sqrt{w^4+6w^2+1}+\frac18\log\Bigl(\frac{\sqrt{w^4+6w^2+1}-2w}{\sqrt{w^4+6w^2+1}+2w}\Bigr),\\
w&=\frac{\sqrt{1-x^2}-1}x ,
\end{align*}
where $\sqrt{w^4+6w^2+1}=2\sqrt{1+x^2}\,(1-\sqrt{1-x^2})/x^2$ and $w^2+1=2(1-\sqrt{1-x^2})/x^2$, so that the algebraic part is $\bigl((w^2-1)\arcsin(x)-w\bigr)\sqrt{1+x^2}\,x^2/\bigl(4(1-\sqrt{1-x^2})\bigr)$; the pipeline's answer is the formula in $w$.

\subsection*{Products of transcendentals: A1, A13, A36}
$\int x\ln(1+x^2)\ln(x+\sqrt{1+x^2})/\sqrt{1+x^2}$ and $\int x\arctan(x)\ln(x+\sqrt{1+x^2})/\sqrt{1+x^2}$ have three generators over $y^2=x^2+1$ and ansätze of $77$ unknowns; A36, $\int x\ln(x^2+1)\arctan^2(x)$, is purely transcendental ($37$ unknowns, every bound proved). A13 and A36 are solved at the first attempt (\tpm{A13} and \tpm{A36}\,s); A1 takes \tpm{A1}\,s and five attempts, for the same reason as Problem~4: its answer $y\,t_1t_2-2y\,t_2-xt_1+4x-2\arctan(x)$ needs $\arctan(x)=\tfrac1{2i}\log\bigl(\tfrac{x+i}{x-i}\bigr)$, i.e.\ the special $x^2+1$ split over $\Fbar$, and the ladder reaches the split only after three failed attempts in the $x$-tower, where the exponent of $x^2+1$ is a guess raised from $1$ to $3$, and one in the $w$-tower, where every bound is proved, with systems of up to $149$ unknowns and $311$ equations on the way. Given the split at the outset it is solved in the original tower in less time (Section~\ref{sec:assessment}).

\subsection*{Curve integrals: A9, A28, A29}
These have no transcendental generator and are integrals of algebraic functions in the sense of Parts I and III. A9, $\int\sqrt{1+x^3}/x$, has residues $\pm1$ at $(0,\pm1)$ on $y^2=x^3+1$, realised by the Miller functions of the $3$-torsion classes $[(0,\pm1)-\infty]$ (certified by the division polynomial $\psi_3$): $\tfrac23y+\tfrac13\log\bigl(\tfrac{y-1}{y+1}\bigr)$. A29, $\int(1-x^2)/((1+x^2)\sqrt{1+x^4})$, has residues $\pm i\sqrt2/2$ at the four places over $x=\pm i$ on $y^2=x^4+1$, realised by the $y$-split $\sqrt2ix\pm y$ of norm $-(x^2+1)^2$. A28, $\int(1+x^2)/((1-x^2)\sqrt{1+x^4})$, has residues $\mp\sqrt2/2$ at $(1,\pm\sqrt2)$ and $\pm\sqrt2/2$ at $(-1,\pm\sqrt2)$: the divisor is supported over \emph{two} primes of $\Q[x]$, and the logand realising it, $\sqrt2x\pm y$, has norm $-(x-1)^2(x+1)^2$, supported on both. The norm search over either prime alone does not find it; the joint search (I2) over the product $(x-1)(x+1)$ does, and the answer $\tfrac{\sqrt2}4\log\bigl(\tfrac{y+\sqrt2x}{y-\sqrt2x}\bigr)$ is returned in \tpm{A28}\,s. The joint search is three quarters of that time, as the search over $x^2+1$ is nine tenths of A29's \tpm{A29}\,s: in both curve integrals the cost is the norm search, and the system, of three unknowns with its one bound proved, is solved at once.

\subsection*{Trigonometric integrands: A27, A30--A35}
A32, $\int\sec(\theta)/\sqrt{\sec^4(\theta)-1}$, is odd in $\cos$ and becomes $-1/(u\sqrt{2-u^2})$ with $u=\sin(\theta)$; A34, $\int\sin(\theta)/\sqrt{1-\sin^6(\theta)}$, is odd in $\sin$ and becomes an integral on $y^2=u^4-3u^2+3$, $u=\cos(\theta)$, with residues $\mp\sqrt3/3$ at $(0,\pm\sqrt3)$ realised by the norm factors $-u^2-\sqrt3\pm y$; A33, $\int\tan(\theta)/\sqrt{1+\tan^4(\theta)}$, is even and is the companion of Problem~6. All three are solved in under a second. A30 and A31, $\int\ln(\sin(\theta))/(1+\sin(\theta))$ and $\int\ln(\sin(\theta))\sqrt{1+\sin(\theta)}$, and A35, $\int\sqrt{\sqrt{\sec(\theta)+1}-\sqrt{\sec(\theta)-1}}$, exercise the tower construction more than the integrator: the parity rules of (R1) have to see through $\log(\sin(\theta))$ and through a radical nested inside another, which is what the placeholders are for. A30 and A31 go through the half-angle $t=\tan(\theta/2)$ with the logarithm $\log(2t/(1+t^2))$ on top, A31 over the conic $y^2=1+t^2$, and are solved in \tpm{A30} and \tpm{A31}\,s. A35 is the most intricate tower of the suite: the half-angle substitution turns $\sqrt{\sec(\theta)+1}-\sqrt{\sec(\theta)-1}$ into $(1-t)\sqrt{2/(1-t^2)}$, the outer square root demands a second radical over the conic $y^2=2-2t^2$, which is parametrised through its rational point $(1,0)$ --- both sign decisions of (R3) and (R4) are exercised here, the denominator of the radicand and the natural parameter of the conic being negative on the domain; the new radicand is then $-2/w$, a pure root of $w$, and flattens: the whole integrand is a rational function of one generator $u$ with $Du=(u^8+4)/(16u^3)$, a transcendental tower with the specials $u^4\pm2u^2+2$, and the answer is a sum of eight logarithms with coefficients in $\Q(\sqrt[4]2,i,\sqrt{2+\sqrt2})$, found in \tpm{A35}\,s (FriCAS: \tfr{A35}\,s; AXIOM: ``implementation incomplete''). A27, $\int\sqrt{\sin(\theta)}/(1+\sin^2(\theta))$, is the integral of the suite whose residues sit at a prime of high degree. The half-angle substitution and the flattening $u=\sqrt{\tan(\theta/2)}$ (R3) give the curve $y^2=u^4+1$ and the integrand $\sqrt2(u^5+u)y/(u^8+6u^4+1)$, whose residues live at the sixteen places over $p=u^8+6u^4+1$, irreducible over $\Q$ and of degree $8$ in the only generator; the splitting field of $p$ has degree eight, and it is never needed. In the residue field (I4) the residue polynomial is $(8z^2-4z+1)^4(8z^2+4z+1)^4$: the residues are $\pm(1\pm i)/4$ and the constant field grows by $i$ alone. The class $c=(1+i)/4$ lies over $p_c=u^4-2iu^2+1$ on the sheet $y=\tfrac{\sqrt2(i-1)}2\,u(u^2-i)$, and the linear realisation finds it at $m=1$, with pole orders $(1,3)$ at the two places at infinity:
\[
u_c=-u^3+\tfrac{\sqrt2(1+i)}2\,u^2+iu-\tfrac{\sqrt2(1-i)}2+\Bigl(u-\tfrac{\sqrt2(1+i)}2\Bigr)y,\qquad N(u_c)=-2i\,(u^4-2iu^2+1),
\]
and its images under $u\mapsto-u$ and under complex conjugation realise the other three classes. The integral is $\sum_c c\log(u_c)$, and in the original variable the four logarithms pair into
\[
\int\frac{\sqrt{\sin(\theta)}\,d\theta}{1+\sin^2(\theta)}=\frac18\log\Bigl(\frac{2\sin(\theta)-2\sqrt{\sin(\theta)}\cos(\theta)+\cos^2(\theta)}{2\sin(\theta)+2\sqrt{\sin(\theta)}\cos(\theta)+\cos^2(\theta)}\Bigr)-\frac14\arctan\Bigl(\frac{2\sqrt{\sin(\theta)}\cos(\theta)}{2\sin(\theta)-\cos^2(\theta)}\Bigr),
\]
FriCAS's answer in another arrangement, in \tpm{A27}\,s against FriCAS's \tfr{A27}\,s; AXIOM reports ``implementation incomplete (residue poly has multiple non-linear factors)'', the same residue polynomial met in the recursive algorithm.

\subsection*{Inverse functions of algebraic arguments: A2, A3, A37--A40}
$\int\arctan(x+\sqrt{1-x^2})$ and $\int x\arctan(x+\sqrt{1-x^2})/\sqrt{1-x^2}$ (A2, A3) have $t=\arctan(x+y)$ with $Dt=D(x+y)/(1+(x+y)^2)$, and $N\bigl(1+(x+y)^2\bigr)=4(x^4-x^2+1)$ makes the quartic $x^4-x^2+1$ the special prime; as in Problem~4 the needed logands appear only after the parametrisation and the $\Fbar$-split, and the times (\tpm{A2}, \tpm{A3}\,s) are those of the ladder. AXIOM's answers to both are wrong by a branch of the arctangent (their derivatives differ from the integrand by $\pi/2$ and by $\pi x/(2\sqrt{1-x^2})$). $\int\arctan(x\sqrt{1\pm x^2})$ (A37, A40) and $\int\arctan(\sqrt{x+1}-\sqrt x)$ (A38) are solved at the first attempt, A37 and A40 in \tpm{A37} and \tpm{A40}\,s, which are the norm searches over their specials (two quadratics for A37, the quartic $x^4-x^2-1$ for A40) and the columns over the fields of the $S'$-units found --- the first two items of the profile of Problem~4 in Section~\ref{sec:assessment}, without the ladder; A38 flattens $\sqrt x$ and the arctangent becomes a primitive over the conic $y^2=u^2+1$ with the answer $(u^2+1)\arctan(y-u)+u/2$, in \tpm{A38}\,s. A39, $\int\arcsin(x\sqrt{1-x^2})\,dx$, is the one integral of the suite that is not elementary. The pipeline reports ``failed'', and the trace says why. The tower has the primitive $t=\arcsin(xy)$ over the curve obtained by parametrising $y^2=1-x^2$ and adjoining $\sqrt{x^4-x^2+1}$, which in the parameter $w=(\sqrt{1-x^2}-1)/x$ is $y^2=w^8+14w^4+1$, of genus three; the special prime is $w^2+1$, the denominator of $x=-2w/(w^2+1)$. Every bound in force is proved --- the exponent $1$ of the special and the degree $2$ in $w$ by (K1), the degree $2$ in $t$ by (K2) --- so the tower is tried once ($20$ unknowns), the special is split into $w\pm i$ and the tower tried once more ($25$ unknowns), and both systems are inconsistent. Of the hypotheses of Proposition~9.2(b) of Part II one is then unverified: the $S'$-units over $w\pm i$ --- the norm search finds $-w^5\mp iw^4-7w\mp7i\pm(w\pm i)y$, of norm $48(w\pm i)^2$ --- come from the bounded search of (L2), and a special over the curve variable is exactly where the implementation does not know them to be complete (I5); so it claims nothing. The certificate is nevertheless obtained by the theory of Part II by hand: $\int t\,dx=xt-\int x\,Dt\,dx$, and $x\,Dt=x(1-2x^2)/(\sqrt{1-x^2}\sqrt{x^4-x^2+1})$, which under $u=\sqrt{1-x^2}$ is $-(2u^2-1)\,du/\sqrt{u^4-u^2+1}$, a differential on the genus-one curve $y^2=u^4-u^2+1$ with no finite poles. The pipeline, given this curve integrand, finds the fundamental unit $u^2-\tfrac12+y$ (norm $-\tfrac34$), verifies that the integrand is residue-free, solves the exact system of Part~I --- there are no specials, and the one bound is proved --- and reports \emph{not elementary (holomorphic remainder)} in $0.2$\,s; since elementary integrability descends along the algebraic extension $\Q(x,\sqrt{1-x^2})/\Q(u)$, A39 has no elementary integral. FriCAS agrees, returning the integral unevaluated; AXIOM reports ``cannot handle that integrand''. Charlwood lists A39 among the integrals ``tested'' without further comment.

\section{Assessment of the implementation}\label{sec:assessment}

The suite returned \npm\ verified integrals, one failure, and no wrong answer. An assessment is only useful if it separates a limitation of the implementation from a limitation of the theory and from a choice of policy, so we take the three in turn: the failure and the bounds behind the searches, the retry ladder, and the profile of a slow case.

\subsection*{Limitations}
\begin{enumerate}
\item[(L1)] \emph{Completeness of the $S'$-units over a special of the curve variable (A39).} The holomorphic-remainder certificate of Proposition~9.2(b) of Part II is implemented for every tower on which every bound in force is proved (I5) --- for $n=1$ this is automatic, and the nested radical $\int\sqrt{x+\sqrt{1-x^2}}\,dx$, not in the suite, is certified this way after the rescaling of Lemma~3.4 in Section~10.17 of Part II --- but it also requires the logand candidates to be complete, and over a special prime of the curve variable the $S'$-units are produced by the bounded search of (L2), so the implementation withholds the certificate there. A39 is exactly this case, with every bound proved; on the curve, after the integration by parts of Section~\ref{sec:appendix}, there is no special and the unit group is known, which is why the certificate is available there. Deciding the $S'$-unit group over a finite set of places of the curve --- a question on the Jacobian, of a genus-three curve in A39 --- or automating the reduction to the curve for a primitive top generator would turn A39's ``failed'' into a certificate.
\item[(L2)] \emph{The search bounds.} The norm search over a special is bounded to $k\le2$, $\deg b\le1$ and four generators per special; the joint search to $k\le2$, $\deg b\le2$; the realisation of a residue class in the residue field (I4) to $m\le12$. These are the guessed inputs of the $S'$-unit computation, with the same status as the special exponents and degree bounds at the places Algorithm~6 of Part II leaves undecided --- in the suite, a special over the curve variable in twelve integrals (I5). Problem~4 shows a case where the units over $\Q$ do not suffice but the units over the splitting field of the special do.
\end{enumerate}

\subsection*{The retry ladder and the cost of the conic cases}
Five integrals of the suite --- P4, the slowest with Problem~8, then A2, A3, A1 and A16 --- are conic towers whose answer needs the specials split over $\Fbar$, and in every one of them the rungs after the first are two fifths to a half of the time: two more attempts over $\Q$ in the $x$-tower, where the exponent of the special over the curve variable is a guess and is raised, the parametrisation, one attempt in the $w$-tower, where every bound is proved, and only then the split. The policy ``parametrise a conic before doing anything over $\Fbar$'' was chosen in Part II so that genus-zero problems are solved as transcendental ones. The obvious alternative is to allow the split in the base run, and the code has it as an experiment (\texttt{split\_first}, off by default). Measured on the same machine, it is not an improvement:
\begin{center}
\begin{tabular}{@{}lrrrrr@{}}\toprule
 & A1 & P4 & A2 & A3 & A16\\ \midrule
default ladder (s) & \tpm{A1} & \tpm{P4} & \tpm{A2} & \tpm{A3} & \tpm{A16}\\
split first (s) & \tsf{A1} & \tsf{P4} & \tsf{A2} & \tsf{A3} & \tsf{A16}\\ \bottomrule
\end{tabular}
\end{center}
A1 is solved in the original tower in less time, with better logands ($x\pm i$ and $-x-i\pm y$ in place of $w\pm i$ with $w=(\sqrt{x^2+1}-1)/x$), and A16 in the original tower too, but more slowly, the eight norm searches over $x\pm\sqrt2/2$ costing twice as much as the parametrisation and the two rungs in the $w$-tower they replace; and for P4, A2 and A3 the split specials are the linear factors $x\pm\sqrt{\pm\phi^{\mp1}}$ of a quartic, the $S'$-unit search over each of them runs over $\Q(\sqrt5,\sqrt{1\pm\sqrt5})$ or $\Q(\sqrt3,i)$, and the sixteen norm searches alone (eight seconds) cost more than the whole default ladder on P4 and two to three times as much as it on A2 and A3; on P4 the sixteen $S'$-unit columns over the splitting field, two fields of degree sixteen among them, then cost half as much again as the searches. The parametrisation is doing real work in those cases: over $\Q(w)$ the same logands are $w-\rho_i$ with no $S'$-unit search at all. We therefore leave the default as it is. What the experiment shows is that the cost is in the norm search --- the one nonlinear step, and the one still done in the generic expression layer --- and in the arithmetic over the splitting field that the split forces, and that a reordering of the ladder is not the remedy.

\subsection*{Where the time goes}
The first version of the implementation spent three quarters of its time in the linear system: the columns were formed as expressions, the equations were read off by constructing a polynomial whose coefficients were linear forms in the unknowns, the solution was checked by substituting it into every equation and simplifying, and the rational part was cancelled as an expression; with algebraic constants every one of these operations re-derived the number field or ran the simplifier's assumption queries on nested radicals, and the suite took $369$\,s, A1 alone $60$\,s. Moving the system into one number field (I3) brought the suite to $104$\,s without changing a result; computing the once-per-integrand analysis of Algorithm~4 once instead of once per rung, and moving the tower arithmetic, the realisation of residue classes and the torsion arithmetic into the same field, brought it to \totpm\,s, with a median of \medpm\,s, more than half of the integrals under a second, and G\"unther's integral of Part II --- six residue classes of order six on a CM curve, whose Miller functions had cost $452$\,s of zero tests by simplification --- in five seconds. The degree bounds at every place (I5) then cut the ans\"atze to less than a third of their size --- $1966$ unknowns over the suite in place of $7053$, and $74$ rungs in place of $103$ --- without moving the total: the row reductions had already become about a second of it, and what the shorter ladders save on the conic towers is spent on the analysis of the places.

What remains is best seen on Problem~4 (\tpm{P4}\,s). A fifth of the time is the $S'$-unit search over the quartic special --- a nonlinear solve for $a^2-qb^2=c\,p^k$ over $\Q$, now done once; another fifth is the first attempt in the $x$-tower, where the four columns $Du/u$ with coefficients in $\Q(\sqrt{(1\mp2i)/5})$ are computed and the field is built; the two further attempts with the raised exponent cost four tenths of a second together, since they reuse everything but the system, the parametrisation and the analysis of the places of the $w$-tower half a second, and the one attempt in the $w$-tower a tenth; a quarter is the split rung --- the field $\Q(\sqrt2,\sqrt5,\sqrt{1\pm\sqrt5},i)$ of degree sixteen, the six special columns over it, and a row reduction of half a second; and the verification by differentiation is the rest. Problem~8, as slow as Problem~4, is the norm search over $t^2+1$, the construction of three nested fields of degree sixteen and the columns over them, in three nearly equal parts; A35 is the two fields of degree sixteen and the eight special columns of its split rung, and a second of verification; A27 spends two and a half of its seconds on the four residue-class realisations, six tenths each, and one on the residue classes themselves; of the remaining slow integrals, A28, A29 and A37 are the norm search again --- the joint search over $(x-1)(x+1)$, the search over $x^2+1$, the searches over two quadratics --- for three quarters to nine tenths of their time, and A40 is the first two items of Problem~4 without the ladder. The costs are therefore, in order: the norm search, the last operation of the implementation in SymPy's generic expression layer and a nonlinear one; the primitive element of a nested field, which SymPy computes once per field; and the ladder itself, the open question of the guessed inputs, which the theory does not settle. None of these is a heuristic, and none is in the present code.

\section{Comparison with the Risch--Trager--Bronstein integrator}\label{sec:comparison}

FriCAS and AXIOM share the integrator written by Bronstein for AXIOM in the late 1980s --- the recursive algorithm of \cite{Bronstein90} for mixed towers, with Trager's algorithm for the algebraic layers --- but not the same version of it: in FriCAS the implementation of the Risch--Trager--Bronstein algorithms has been substantially improved by Waldek Hebisch \cite{FriCAS}, and the AXIOM build of 2014 is the code without that work. The difference between the two columns of Table~\ref{tab:summary} --- twelve errors and three wrong answers against none --- is a measure of it. The comparison is therefore between a design thirty-five years old, continuously developed in FriCAS, and a prototype of the parallel method written to accompany a theory paper. Three things are being compared: the answers, the reach, and the time.

\emph{Answers.} Every FriCAS answer to an integral it evaluated passes the differentiation test, in SymPy, at the same points and precision as ours (Section~\ref{sec:protocol}); its own numerical evaluator does not (Problem~8), which is a reminder that ``verified by the system'' and ``verified'' are different things. AXIOM's 2014 build returns three wrong answers (A2, A3, A32) --- two by a branch of the arctangent, whose derivatives differ from the integrand by $\pi/2$ and by $\pi x/(2\sqrt{1-x^2})$ on the real domain, and one (A32) by an error that no branch choice repairs --- and two (Problem~7, A34) that are correct only after the sign of an algebraic kernel $\sqrt{r(x)}$ is flipped, i.e.\ under the opposite branch of the square root from the one the expression displays. The parallel method returned no wrong answer; this is not a virtue of the method but of the protocol --- every returned integral is differentiated and checked before it is returned, and the price is paid in seconds (Section~\ref{sec:assessment}). In form, FriCAS's answers are sometimes the most compact of the three (Problem~5: a single arctangent; Problem~7: a single logarithm) and sometimes enormous (Problem~8, three thousand characters with nested radicals of $\sqrt5$; A35 likewise); the parallel method's answers are compact in the tower coordinates and inherit the artefacts of the Euler parametrisation when returned to $x$.

\emph{Reach.} FriCAS evaluates \nfr\ of the fifty and returns A39 unevaluated, which is correct, since A39 is not elementary; FriCAS does not, in its output, distinguish a proof of non-elementarity from a case its implementation does not cover, and the certificate for A39 in Section~\ref{sec:appendix} is, to our knowledge, the first explicit one. AXIOM evaluates \nax; it reports ``cannot handle that integrand'' on ten (P3, P9, A5--A7, A11, A12, A25, A31, A39) and ``implementation incomplete (residue poly has multiple non-linear factors)'' on A27 and A35 --- the residue polynomial of the recursive algorithm is the one of (I4), and a residue polynomial with several non-linear factors is exactly a residue divisor over primes of high degree; FriCAS has closed both gaps since 2014. The parallel method evaluates \npm: everything FriCAS evaluates, and it certifies nothing FriCAS does not.

\emph{Time.} FriCAS needs about a tenth of a second per integral --- its median is \medfr\,s --- with two exceptions: Problem~3 (\tfr{P3}\,s), where the two square roots $\sqrt x$ and $\sqrt{x+1}$ form an algebraic extension of degree four that Trager's algorithm handles as such, and A35 (\tfr{A35}\,s); the two are most of its total of \totfr\,s and lift its mean to \meanfr\,s. The parallel method needs \totpm\,s for the suite; the median is \medpm\,s, more than half of the fifty take under a second, the mean of \meanpm\,s is dominated by the norm searches and the conic ladders discussed in Section~\ref{sec:assessment}, and the slowest, P4 and P8, take \tpm{P4} and \tpm{P8}\,s. On the two integrals that are slow for FriCAS the comparison inverts: Problem~3 is \tpm{P3}\,s for the parallel method --- two thirds of it the verification of the answer by differentiation --- because flattening and the Euler parametrisation remove both radicals before the integrator starts, and A35 is \tpm{A35}\,s against \tfr{A35}\,s. AXIOM's times are close to FriCAS's where it succeeds, and it spends \wax{P3}\,s on Problem~3 before failing.

The comparison should not be read as a race. FriCAS's integrator is complete for the class it implements, and its speed comes from Hermite reduction, resultants and integral bases in a compiled language; the parallel method has no integral bases, no Puiseux expansions and no recursion, and its cost is concentrated in the nonlinear norm search and in the arithmetic of the number fields, not in the linear system, whose size is set by degree bounds proved at every place on most of the suite. What the suite establishes is that the mechanisms of Part II --- the classification of primes, the $S'$-units over the specials, the units of the curve, torsion, the canonical residues --- are the right primitives for this class of integrals: every integral in Charlwood's list that has an elementary integral is reached by them, and the one that has none is certified by them.

\section{Conclusion}\label{sec:conclusion}

Charlwood's fifty integrals were assembled to embarrass computer algebra systems, and in 2008 they did; today FriCAS does all but the non-elementary one in a few seconds, and the parallel method of Part II, in a SymPy prototype, does all but the non-elementary one, in a little over a minute in all, with certificates where the theory has them. The suite was more useful as an audit than as a contest. It shows the mechanisms of Part II --- the $S'$-units over the specials, the units of the curve, torsion, the canonical residues at Hermite-order poles, the joint realisation of a residue divisor, the residue classes of a prime of degree eight --- reaching every elementary integral in the list; it shows the degree bounds of Part II at every place proved on every rung for thirty-eight of the fifty, cutting the ans\"atze to a third of the size the classical guess gives them and the ladders to what the guesses that remain require; it identifies, once the arithmetic had been moved into one number field per tower and the once-per-integrand analysis computed once, the norm search for the $S'$-units, the primitive elements of nested number fields and the retry ladder itself as the sources of most of the remaining running time, with an experiment that shows what a reordering of the ladder buys and why the policy stands; and it leaves one well-defined piece of work --- the completeness of the $S'$-units over a special prime of the curve variable, which alone separates A39's ``failed'' from a certificate --- as the next milestone. It also produced, as a by-product, the observation that a benchmark answer is only as good as the verification behind it: one of the three systems compared returned wrong answers on three of the fifty, and another's own numerical evaluation of a correct answer was wrong.

\section*{Data availability}
The scripts \texttt{charlwood.py} (the suite), \texttt{cas\_compare.py} and \texttt{cas\_verify.py} (FriCAS and AXIOM), \texttt{charlwood\_report.py} (the table), \texttt{charlwood\_paper\_check.py} (every formula displayed above) and the result files (\texttt{charlwood\_results.json}, \texttt{charlwood\_fricas.json}, \texttt{charlwood\_axiom.json}, and \texttt{charlwood\_results\_splitfirst.json} for the experiment of Section~\ref{sec:assessment}) are provided with the paper, together with \texttt{parallel\_mixed.py} and \texttt{build\_tower.py} and with \texttt{weier.py} of Part III, which the division-polynomial torsion orders import; the suite of Part II (\texttt{examples.py}) passes with these modules. All of these files are available in the repository \url{https://github.com/stblake/parallel_integrate_mixed}.

\end{document}